\documentclass[prl, twocolumn, superscriptaddress, letterpaper, showpacs, amsfonts]{revtex4}
\usepackage{epsfig}

\begin{document}
\bibliographystyle{apsrev}

\title{Thermodynamics of C incorporation on Si(100) from 
{\em ab initio} calculations}

\author{I. N. Remediakis}
\affiliation{Department of Physics and Division of Engineering and 
Applied Sciences, Harvard University, Cambridge MA 02138}
\affiliation{Department of Physics, University of Crete, 71003
Heraklion, Crete, Greece}

\author{Efthimios Kaxiras}
\affiliation{Department of Physics and Division of Engineering and 
Applied Sciences, Harvard University, Cambridge MA 02138}
\affiliation{Institute of Electronic Structure and Laser, Foundation
for Research and Technology-Hellas (FORTH), Heraklion, Crete, Greece}
 
\author{P. C. Kelires}
\affiliation{Department of Physics, University of Crete, 71003
Heraklion, Crete, Greece}
\affiliation{Institute of Electronic Structure and Laser, Foundation
for Research and Technology-Hellas (FORTH), Heraklion, Crete, Greece}

\date{\today}

\begin{abstract}
  We study the thermodynamics of C incorporation on Si(100), a system
  where strain and chemical effects are both important. Our analysis
  is based on first-principles atomistic calculations to obtain the
  important lowest energy structures, and a classical effective
  Hamiltonian which is employed to represent the long-range
  strain effects and incorporate the thermodynamic aspects.  We
  determine the equilibrium phase diagram in temperature and C
  chemical potential, which allows us to predict the mesoscopic
  structure of the system that should be observed under experimentally
  relevant conditions.
\end{abstract}

%
%
%
%

\maketitle

Carbon-enriched silicon systems are the focus of current interest as
candidates for a material with tailored electronic properties,
which is compatible with well-established silicon technology. The
tetravalent nature and large band gap in the diamond structure make
carbon an ideal candidate for incorporation in Si.  However, the
solubility of C in Si under thermodynamic equilibrium is extremely low
($\approx 10^{-5}$) due to the huge mismatch in bond length (35\%) and
bond energy (60\%) between C and Si. Non-equilibrium methods, such as
molecular beam epitaxy (MBE), that exploit the higher atomic mobility
on surfaces, can be used to overcome this obstacle and enhance
solubility \cite{rucker94}. As predicted theoretically by
Tersoff\cite{tersoff95}, C solubility is enhanced by several orders of
magnitude near the Si(100) surface, especially in subsurface
layers. Osten {\em et al.} \cite{osten95} confirmed experimentally
this prediction and observed that C atoms diffuse to subsurface layers
above a certain temperature. This finding opened new possibilities for
growth of C-rich metastable structures.

The enhanced solubility near the surface has important
consequences. The large tensile strain associated with C incorporation
in Si has proven a very powerful tool in device engineering: a small
amount of C can compensate the Ge-induced tension in pseudomorphic
SiGe layers \cite{eberl92,osten94,kelires95} , and can also suppress
dopant outdiffusion. This idea was recently implemented in a novel
heterojunction bipolar transistor \cite{osten_trans}.  Another
interesting effect\cite{butz98,miki97} produced by C incorporation on
the Si(100) surface is an unusual change of the surface periodicity
after deposition of even a small amount of C ($\approx \frac{1}{8}$of
a mono-layer (ML)): the well-known $c(2\times 4)$ or $p(2\times 2)$
reconstructions of the pure Si surface change to a $c(4 \times 4)$
pattern. This is clearly visible in several LEED experiments after
ethylene exposure \cite{shek98,takaoka96}, or MBE \cite{butz98}.

The microscopic features of C incorporation in Si are rather well
understood. Previous work by the authors
\cite{kelires97,kelires98,kelires98_2} revealed an oscillatory C
profile driven by the competition between two factors: the tendency of
C atoms to occupy favorable sites which are determined by the
reconstruction strain field, and the preferential arrangement of C
atoms at certain distances which minimizes the lattice elastic
energy. The profile is characterized by enhancement of C content in
the first and third layer, depletion in the second and an exponential
reduction from the fourth layer and beyond. The favorable C sites are
in the third layer between the surface dimers, being under compressive
stress \cite{kelires89,tersoff95} and thus suitable for the
smaller-sized C atom. The preferential arrangement of two C atoms in
the surface layers is at a third nearest-neighbor position, which is
also the lowest-energy configuration in the bulk \cite{rucker94}.  The
interaction of C atoms in first nearest neighbor position is highly
repulsive \cite{kelires97}. Moreover, a recent experiment
\cite{butz98} showed that C-C dimers are quite rare in the annealed
C-Si(100) surface.

\begin{figure}
\label{fig:confs}
\begin{center}
\epsfxsize=0.55\linewidth\centerline{\epsfbox{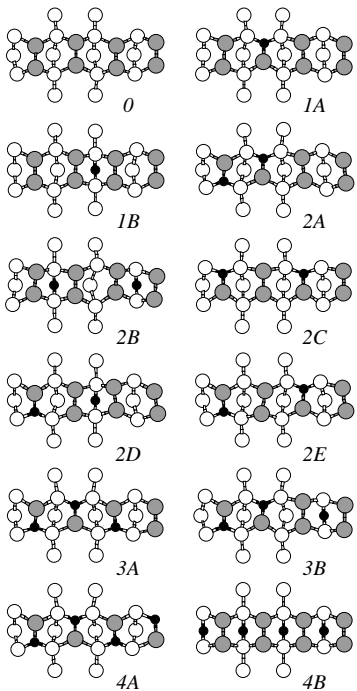}}
\end{center}
\caption{ Top views of the relaxed geometries of the 12 $c(4\times4)$ cells
  we considered.  Si surface atoms are shown as grey circles, Si atoms
  in 2nd and 3rd layers as smaller, white circles, and C atoms as
  black circles.}
\end{figure}

On the other hand, a mesoscopic picture of the surface structure that
would link the atomistic features to long-range strain effects, and
predict structures for different growth conditions more relevant to
experiment, is lacking. Here, we present an approach that closes this
gap and is able to determine the equilibrium surface phase diagram. It
is based on first-principles atomistic studies, incorporating
microscopic strain and polarity effects, which are expected to have a
dominant role in the understanding of C incorporation in the host Si
lattice. The results are linked to a classical effective Hamiltonian
in a Monte Carlo scheme that incorporates strain into the thermodynamic
aspects of the problem.

We considered first all possible atomistic configurations likely to
have low energies. Guided by previous work,
\cite{tersoff95,kelires97,kelires98,leifeld99} we can establish a set
of rules which should be obeyed by low energy configurations of the
C-Si(100) system, namely:

(i) The surface has $c(4\times 4)$ periodicity.

(ii) C substitutes Si atoms only in the first (surface) or in the third layer.

(iii) There are no C-C nearest neighbors, or second nearest neighbors.

(iv) Third layer C exists only at sites between pairs of surface dimers.

\noindent We constructed every possible structure of the C-Si(100) system,
consistent with these rules, with 0, 1, 2, 3 and 4 C atoms per
$c(4\times4)$ unit cell. These are shown in Fig. 1. The configurations
are named $nX$, $n$ being the number of C atoms in the $c(4\times4)$ unit
cell and $X$ an index to distinguish structures of the same $n$.

To study the energetics and the geometrical features of these
configurations, we performed {\em ab initio} DFT/LDA/Pseudopotential
calculations \cite{method}. In Table I, we give the energy $E(nX)$ for
each configuration $nX$, relative to the pure Si(100) surface
(configuration 0). These energies are defined by:
\begin{equation}
\label{eq:ec}
E(nX)=E_{tot}(nX) - n (\mu_C - \mu_{Si}) - E_{tot}(0),
\end{equation}
where $E_{tot}(nX)$ is the calculated total energy of configuration
$nX$, and $\mu_C, \mu_{Si}$ are the chemical potentials for C and Si
atoms. For Si, the chemical potential is taken to be the energy of the
bulk, since this is the natural reservoir for Si atoms due to the
presence of steps, terraces, and other surface defects. As a starting
point, $\mu_C$ is taken to be the energy of a C atom in bulk
diamond. We discuss its variation below.

For low concentration of C, one atom per unit cell, which corresponds
to $\frac{1}{8}$ ML coverage, the structure with subsurface C ($1B$)
has lower energy than the one with a Si-C dimer ($1A$) on the
surface. This result is in agreement with experiment: X-ray spectra
\cite{shek98} indicate that for low C deposition, there is almost no C
on the surface. A single C atom at a third-layer site relieves the
compression due to the surface reconstruction and makes four Si-C
bonds while a single C atom on the surface, as part of a dimer, allows
the formation of only three SiC bonds.
Passing to configurations with more than one subsurface C atom per
unit cell, we find that the energy of configuration $2B$ is
considerably higher than that of $2A$, $2C$, and $2E$. Apparently, the
presence of two subsurface C atoms per unit cell produces a large
distortion of the surrounding Si-Si bonds. Thus, for two C atoms in
the unit cell ($\frac{1}{4}$ ML coverage), the best situation is to have both C
atoms on the surface.
For three C atoms per unit cell ($\frac{3}{8}$ ML coverage), we
observe an equivalence of surface/subsurface sites: configurations
$3A$ and $3B$ have almost the same energy. For the case of four C
atoms ($\frac{1}{2}$ ML coverage), there is just one structure ($4A$)
consistent with the rules we discussed before. For illustration
purposes, we include another configuration, $4B$, containing C atoms
in second neighbor positions, which violates rule (iii); its very high
energy can be considered as a justification for these rules.  We do
not go beyond $\frac{1}{2}$ ML coverage because experiments show that
this results in disordered structures.

\begin{table}
\begin{center}
\begin{tabular}{|c|c||c|c||c|c||c|c|} \hline
 $nX$ & $E(nX)$ & $nX$ & $E(nX)$ & $nX$ & $E(nX)$ & $nX$ & $E(nX)$ \\ \hline
$1A$ & 0.19 & $2A$ &  0.35 & $3A$ & 1.25 & $4A$ & 2.31 \\ 
$1B$ & 0.08 & $2B$ &  0.81 & $3B$ & 1.28 & $4B$ & 10.39 \\
     &      & $2C$ &  1.06 &      &      &      &       \\ 
     &      & $2D$ &  0.49 &      &      &      &        \\ 
     &      & $2E$ &  0.35 &      &      &      &  \\ \hline
\end{tabular}
\end{center}

\caption{ Relative energies for the 12 configurations of 
Fig. 1, according to Eq. (1) in eV per $c(4\times4)$ unit cell. 
$\mu_C$ is taken as the energy of a C atom in diamond. \cite{chempot}}
\end{table}

To compare configurations with different C content, we use Eq. (1).
This implies that the surface is in equilibrium with a
reservoir of C atoms characterized by $\mu_C$. It is generally
accepted that C forms small clusters, whose cohesive energies are in
the range between $-5.5$ and $-7$ eV \cite{jones99}. In Fig. 2(a), we
show the energies of configurations $0$, $1B$, $2A$, $3A$ and $4A$ as
a function of $\mu_C$. The vertical faint lines denote the transition
points at which the preferred structure changes. We observe that the
lowest energy structure depends strongly on $\mu_C$ and hence on the
conditions of C deposition, making it difficult to predict what the
actual structure of the system will be. This implies that the
equilibrium surface structure might be composed of different
configurations at the atomistic level and, consequently, a
larger-scale mesoscopic picture is needed. In this picture, the
elastic interactions in the boundaries between regions of different C
content or atomistic structure will play an important role.

\begin{figure}
\epsfxsize=0.9\linewidth\centerline{\epsfbox{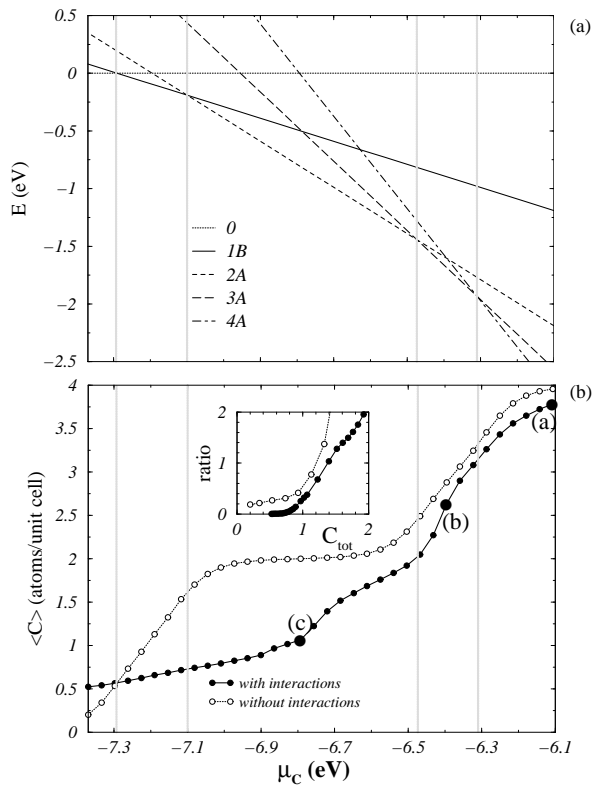}}
\caption{ (a) The energy of the preferred structure for 1,2,3 and 4 C
atoms per unit cell as a function of $\mu_C$, from Eq. (1).  (b) The
average C content of the system as a function of C chemical potential,
at 850K, with (filled symbols) and without (open symbols) the
elastic strain interactions. The inset shows the ratio of surface to
subsurface C atoms at the same temperature, as a function of the total
C content in the unit cell. (points marked with bigger symbols
correspond to the cases shown in Fig. 3).}
\end{figure}

To obtain such a mesoscopic description and study its thermodynamics,
we use an effective Hamiltonian within a Monte Carlo scheme. We first
construct a classical, 12-state generalized Potts model. We assume
that the surface consists of $c(4\times4)$ cells, each of which can
exist in any of the lowest energy configurations discussed before. The
Hamiltonian is
\begin{equation}
\label{eq:ham}
\mathcal{H}_{eff}=\sum_{i=1}^{N}[ E(c_i)+\frac{1}{2} \sum_{j} \Delta E(c_i, c_j)],
\end{equation}
where the summation on $i$ runs over all $N$ cells, and the summation
on $j$ over the 8 nearest neighbors of $i$ (the factor of one half
takes into account double-counting of the interaction); $c_i$ is the
configuration in cell $i$, and $\Delta E(c_i, c_j)$ is the interaction
energy between the neighboring cells $c_i$ and $c_j$ ($\Delta E(c_i,
c_i)\equiv 0$).  In order to obtain these interaction energies, we
calculated the total energies of all 144 $c(12\times 12)$ cells
consisting of a cell $c_i$ surrounded by 8 $c_j$ cells. These were
obtained using Tersoff's empirical potential \cite{tersoff89}, with a
Monte Carlo relaxation method. The key point here is that we are
interested in elastic strain interactions between different cells
rather than any specific features of their structure; for this type of
interaction {\em between cells} the classical potential employed gives
very reasonable results as established by previous studies of similar
systems \cite{kelires95,kelires98}. Having determined
$\mathcal{H}_{eff}$, we performed an equilibrium Monte Carlo study for
different values of the temperature and $\mu_C$. We used a $70 \times
70$ grid of cells, which was sufficient to get converged averages. For
each value of $\mu_C$ the system was started from a random
configuration at high temperature (1200K) and then was cooled down to
room temperature.

In Fig. 2(b) we plot the average total C content of the system as a
function of $\mu_C$. The zero for $\mu_C$ is taken to be the energy of
an isolated C atom. For comparison, we plot the same quantity when the
elastic strain interactions, $\Delta E(c_i,c_j)$, are set equal to
zero. In both cases the average C content of the system increases with
increasing $\mu_C$. The difference between the two curves shows the
importance of elastic interactions, which, being of the order of
0.1-0.2 eV/ $c(4\times4)$ cell, are more important than the total
energy differences in the low $\mu_C$ region. This effect weakens as
$\mu_C$ increases.

Another important consideration, with experimentally observable
consequences, is the relative amount of surface versus subsurface C
atoms as a function of the total C content $C_{tot}$ in the unit cell,
shown in the inset of Fig. 2(b). For larger total C coverage the ratio
increases monotonically, approaching infinity as $C_{tot}\rightarrow
4$. Again, we give the case with $\Delta E (c_i,c_j)= 0$ for
comparison.  There is a critical coverage of 1.39 C atoms per unit
cell, or 0.17 ML, beyond which C prefers mostly surface sites. This
critical coverage has a weak temperature dependence: it has a maximum
around 850K, and then falls monotonically down to roughly 0.16 ML at
300K and 1100K. The increase of the surface C with increasing total C
is not unexpected: although third layer compressed sites
seem ideal for the smaller C atom, a large amount of subsurface C
would cause large strain and raise the total energy dramatically, as
exemplified by the case of configuration $4B$, which we discussed
earlier. From Fig. 2(b) we see that the transition from mostly
subsurface to mostly surface C content does not coincide with the
change from 1 to 2 C atoms per cell, as suggested by our {\em ab
initio} results, indicating that {\em the elastic strain interactions
play a dominant role in the structure of the surface}.

Representative snapshots of the C distribution from the Monte Carlo
run at 850K are shown in Fig. 3. The three cases correspond to the
points shown in Fig. 2(a). An impressive long range order is revealed
for the subsurface C (right column of Fig. 3). For low $\mu_C$,
Fig. 3(c), this order appears in the form of alternating rows of
similar C content. These consist mostly of unit cells corresponding to
$0$ and $2B$ structures. Although structure $2B$ has relatively high
energy, as shown in Table I, a structure of alternating $0$ and $2B$
cells seems to be preferable, giving C the chance to take advantage of
the desirable third layer compressed sites. The line-pattern is a
result of the strong interactions in that range of $\mu_C$, which
forces the $c(4\times4)$ cells to have the maximum number of different
neighbors, 6 for the row pattern and 4 for a chess pattern. The latter
pattern is observed for higher $\mu_C$, Fig. 3(b), and is typical for
the range of $\mu_C$ and temperature corresponding to 1-2 C atoms per
unit cell on average (see also Fig 2(b)). In this case, the
self-energies of the different configurations dominate over the
elastic strain interactions. In the specific example shown in
Fig. 3(b), the structure consists mostly of $3A$ and $2B$ cells, at a
ratio 2:1. This suggests that for conditions corresponding to this
range of $\mu_C$, an alternating pattern of surface/subsurface C is
preferable. As $\mu_C$ increases the energies of the cells become
large enough so that the elastic interactions are not that important,
and thus the system does not have much to gain by
self-organizing. This results in the random pattern shown in
Fig. 3(a), which consists mainly of $4A$ cells with a few $3B$ cells
around them. Again, of the two configurations with three C atoms, the
one with some subsurface C is preferred when the dominant
configuration has all its C atoms on the surface.

Finally, we note that while subsurface C shows this interesting
ordering behavior, the surface C seems to be randomly distributed as
the patterns of the middle column of Fig. 3 indicate.  The ordered
structure appearing in the total distribution of C (left column of
Fig. 3) is caused by an ordered third-layer C configuration; this is
best seen in Fig. 3(c). This idea was implicitly suggested by previous
experimental work \cite{shek98,butz98}. From our first-principles
calculations, we found that the geometrical features of the Si-Si
dimers in all 12 cells are essentially the same. Surface C is
invisible, or shows up as a missing Si atom, in STM experiments
\cite{leifeld99}. This implies that one cannot distinguish, using
standard microscopy, between pure Si(100) or Si(100) with vacancies,
and a configuration with C in the third layer; the only criterion is
the long-range order and the change of the reconstruction.

\begin{figure}
\label{fig:ctot}
\begin{center}
\epsfxsize=0.9\linewidth\epsfbox{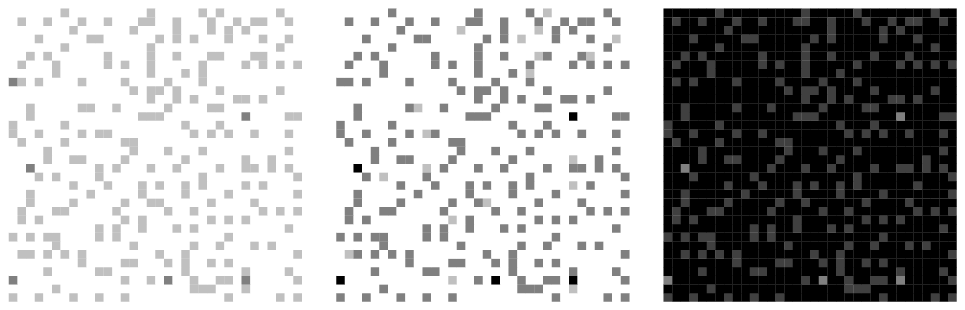}(a)

\epsfxsize=0.9\linewidth\epsfbox{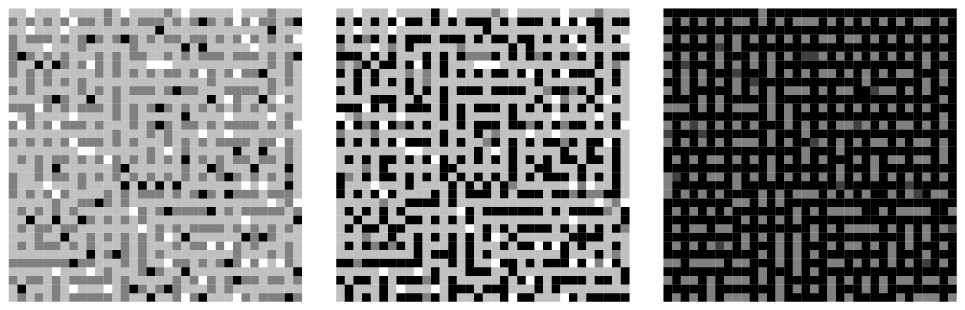}(b)

\epsfxsize=0.9\linewidth\epsfbox{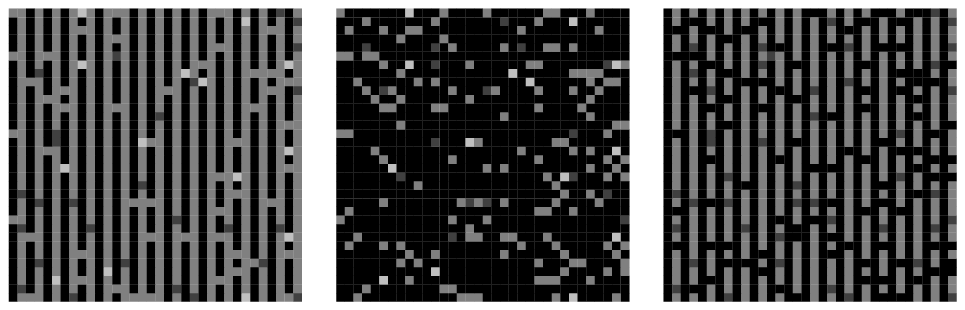}(c)
\begin{tabular}{p{2.65cm}p{2.5cm}p{2.66cm}} 
 ~~Total & Surface & Subsurface
\end{tabular}
\end{center}
\caption{ Density plot of the C distribution on the surface at T=850K;
(a) $\mu_C=-6.1$, (b) $\mu_C=-6.4$eV, (c) $\mu_C=-6.8$ eV for
(c). White corresponds to 4 C atoms, black to 0 C atoms. Each panel
represents an area of 38$\times$38 nm$^2$ and is $\frac{1}{4}$ of our
simulation cell.}
\end{figure}


The work of INR and PCK is supported by a program of the Greek General
Secretariat for Research and Technology. EK acknowledges the
hospitality of IESL/FORTH where part of this work was completed.



\begin{thebibliography}{99}
\bibitem{method} We use a periodic slab of 7 layers terminated by H on
one side with 6 layers allowed to relax. We use a plane wave basis
with a cutoff kinetic energy of 36 Ry and the $\Gamma$ point of the
Brillouin zone for reciprocal space sampling.
Atomic positions are relaxed until the magnitude of the calculated
Hellman-Feynman forces is smaller than 0.001 Ry/a.u. For each
configuration, several different starting geometries were considered
to ensure that the correct minimum energy structure is found.
\bibitem{chempot} To obtain the chemical potentials, we calculate the
energies of C and Si in the diamond structure with the same kinetic energy
cutoff and the corresponding k-point grid spacing for the bulk. We use
the experimental cohesive energy of diamond, namely 7.37 eV/atom.
\end{thebibliography}
\end{document}